# Photometry and Long-Slit Spectroscopy of Split Comet C/2019 Y4 (ATLAS)

Oleksandra Ivanova,[1,2,3*], Igor Luk'yanyk,[3] Dušan Tomko,[1] Alexei Moiseev[4]

[1] *Astronomical Institute of the Slovak Academy of Sciences, SK-05960 Tatranská Lomnica, Slovak Republic*
[2] *Main Astronomical Observatory of the National Academy of Sciences of Ukraine, 27 Zabolotnoho Str., 03143 Kyiv, Ukraine*
[3] *Taras Shevchenko National University of Kyiv, Astronomical Observatory, 3 Observatorna Str., 04053 Kyiv, Ukraine*
[4] *Special Astrophysical Observatory of the Russian Academy of Sciences, 369167 Nizhny Arkhyz, Russia*

[*] Corresponding author's e-mail: oivanova@ta3.sk

**ABSTRACT**

We present an analysis of the photometric and spectroscopic observations of the split comet C/2019 Y4 (ATLAS). Observations were carried out on the 14$^{th}$ and 16$^{th}$ of April 2020 when the heliocentric distances of the comet were 1.212 and 1.174 au, its geocentric distances 0.998 and 0.991 au, and the phase angle 52.9° and 54.5°, respectively. The comet was observed with the 6-m BTA telescope of the Special Astrophysical Observatory (Russia) with the SCORPIO-2 multi-mode focal reducer. The narrow-band *BC* and *RC* cometary filters in the continuum were used. We identified numerous emissions of the CN, $C_2$, $C_3$, and $NH_2$ molecules within the range of 3750 — 7100 Å. The $C_2$/CN and $C_3$/CN production rate ratios coincide with those of typical comets. Four fragments belonging to the coma were detected in both observational runs. We compared and analyzed temporal variations of the visual magnitudes, gas productivity, and dust colour. Based on our dynamical investigation of the orbits of comets C/1844 Y1 (Great comet) and C/2019 Y4 (ATLAS), we can claim that, with high probability, two comets do not have a common progenitor.

1.   **Introduction**

About 40 comets after their splitting have been observed in the last 150 years (Boehnhardt, 2004). Analysis of the post-splitting dynamics of cometary fragments showed that the splitting of the long-period comets can happen only at large heliocentric distances (over 50 au), while the nucleus of short-period comets can split wherever along the orbit. Partial fragmentation of the cometary nucleus or its complete disintegration allows us to study the internal structure and composition of the nucleus. At the moment, we do not know whether there are cometesimals that have been created in the early formation history of the Solar system or they are fragments from a considerably transformed surface crust of the parent body. As a decay of nuclei into pieces is not a common property of comets, it is observed very rarely. Comet splitting is a significant process of the nucleus mass loss and therefore may play a key role in the cometary nucleus evolution.

There are two main types of comet splitting (A and B, Boehnhardt, 2004) with two different model interpretations: "(A) – the split comet has a few (usually two) components. The primary fragment is the one that remains "permanent"; the secondary can be minor, short-lived, or persistent for a longer time (years to centuries). The primary is considered to be identical to the original nucleus (the parent body), while the secondary represents a smaller piece that is broken off the nucleus (typically 10–100 m in size). Type A splitting events can recur in the same object; (B) – the split comet has many (more than 10) components that could arise from a single or a short sequence of fragmentation events. The fragments are short-lived (possibly of small size), and no primary component can be identified. Tertiary fragmentation of secondaries is occasionally observed." Most components of a split comet "disappear". In other words, on the time intervals from hours to years, the components become undetectable even by the largest telescopes, while the main component is only one to "survive" over a long time period.

According to Sekanina (1997), a decay of cometary nuclei can occur in two ways: i) a cometary cortex separates from the nucleus that rapidly rotates, or ii) a rapidly rotating nucleus, in which the tensile strength and density are rather small, decays. Comet 73P/Schwassman-Wachmann 3 exemplifies the first mechanism described. Presumably, this comet had a very inhomogeneous inner nucleus structure and dense surface crust. Owing to a great flare, several fragments separated from the nucleus surface.

Under the assumption that comets are an assembly of several planetesimals, they most probably divide into individual fragments (secondary nuclei) along the boundaries of planetesimals - regions of a weaker connection in the nucleus. The difference observed in the activity of fragments can be due to physical and chemical differences of cometesimals. Evaporation rates can vary due to physical differences, such as an inhomogeneous density distribution. A certain inhomogeneity of the nucleus physical structure can be accompanied by the heterogeneity of its chemical composition. Individual inclusions, i.e., chemical differences of cometesimals are also probable. Any optical emission differences give proofs of inhomogeneity of the primary nucleus.

Comparison of the unexcitable components (organic and inorganic dust) in different comets also helps to reveal the similarity and difference of comets. The infrared radiation of dust in the continuum and their comparison with the properties of the interstellar dust can reveal differences in the composition of cometary nuclei if they are formed from the interstellar dust. Dust contributes significantly to the formation of crusts and the screening of the icy surface of the nucleus, and its performance is mainly due to gas and dust emissions. Therefore, the relative gas/dust composition in different comets can also indicate the homogeneity of both their nuclei and the regions, in which the comet nuclei were formed.

The long-period comet C/2019 Y4 (ATLAS) (hereafter, 2019Y4) was found by the Asteroid Terrestrial-Impact Last Alert System (ATLAS) at Mauna Loa, Hawaii, on UT 2019 December 28.61 (Staff, 2020 https://minorplanetcenter.net/mpec/K20/K20AB2.html). The existing orbital solution suggests that the comet orbits in an elliptical trajectory around the Sun having the eccentricity $e$=0.99924, the inclination $i$=45.4°, the perihelion distance $q$=0.2528 au, and an orbital period of 6011.43±44.33 yr (https://ssd.jpl.nasa.gov/sbdb.cgi#top). The comet passed the perihelion on May 31, 2020. Similarities were noted between the orbital elements of 2019Y4 and the Great comet of 1844 (C/1844 Y1) suggesting that 2019Y4 is a fragment of the same parent body, split about five thousand years ago (Green, 2020; Hui and Ye, 2020). L. Denneau (MPEC 2020-A112: comet C/2019 Y4 (ATLAS)) was the first who identified cometary signs in a new object; further observations in the following nights confirmed the presence of a coma and a noticeably increasing comet tail (Green, 2020).

At the time of the discovery, the comet's apparent brightness was 19.6$^m$ (Green, 2020). The brightness of the comet from the beginning of February to almost the third decade of March 2020 increased from 17.2 to 14.7$^m$ (Green, 2020). It was expected that by the time of the passage of the perihelion, the visible brightness of the comet should have reached 3 — 4 magnitudes. But around March 22, 2020, the comet started to disintegrate. In April 2020, the Astronomer's Telegram (Ye, Hui, 2020a,b; Steel et al., 2020; Lin et al., 2020) reported the possible disintegration of comet 2019Y4. The comet has fragmented into at least 4 pieces (Ye and Hui, 2020b). The comet observation with the Hubble Space Telescope showed near 30 fragments on April 20 and 25 pieces on April 23 (Andreoli et al., 2020). The Solar Orbiter flew through the ion tail of comet 2019Y4 between May 31 and June 1 and was passing through the dust tail on June 6 (Jones et al., 2020).

The disintegration of the nucleus may be due to the gas outburst caused by the rise of the comet's centrifugal force; the estimated velocity of the fragment expansion is of the order of several m/s (<10 m/s) (Green, 2020; Hui and Ye, 2020). Note that in Boehnhardt (2004) it was shown that the expansion velocity of fragments of decaying comets is within 0.1<$V_{sep}$<15 m/s. This decay mechanism is also indicated by a significant radial non-gravitational effect of the comet's heliocentric motion revealed by astrometric observations in January-April 2020 (Hui and Ye, 2020). It is believed that such fragmentation is typical of the sungrazing comets of the Kreutz group, i.e., the comets close to the Sun. According to some estimates, the diameter of the nucleus of comet 2019Y4 before its decay was 120 m (Hui and Ye, 2020). Tisserand's parameter of comet 2019Y4 relative to Jupiter is $Tj$=0.454 (https://ssd.jpl.nasa.gov/sbdb.cgi#top) which is typical for comet-like orbits.

In this paper, we describe and analyze our observations of split comet 2019Y4. Details of the photometric and spectroscopic observations of the comet and data reduction are given in Section 2. The analysis of spectroscopic and photometric observations is presented in Sections 3 and 4, respectively. The dynamical relationship between comets C/1844 Y1 (Great comet) and 2019Y4 is investigated in Section 5. The summary is in Section 6.

## 2. Observations and Data Reduction

### 2.1. Instruments and General Features

Comet 2019Y4 was observed on April 14 and April 16, 2020, with the 6-m BTA telescope of the Special Astrophysical Observatory (Russia). The comet was at the heliocentric distance of 1.212 and 1.174 au, at the geocentric distance of 0.998 and 0.991 au, and at the phase angle of 52.9° and 54.5°, respectively. Observations were carried out 47 and 45 days before the perihelion passage of the comet (May 31, 2020). The multi-mode focal reducer SCORPIO-2 (Spectral Camera with Optical Reducer for Photometrical and Interferometrical Observations, Afanasiev & Moiseev, 2011) mounted in the primary focus of the telescope was used both in the photometric and spectroscopic modes (Afanasiev and Moiseev, 2011). The CCD detector E2V CCD261-84 of 2048×4104 pixels and the 15×15-μm pixel size was used. The full field of view of the CCD chip is 6.8×6.8' with an image scale of 0.4″/px. To increase the S/N ratio of the observed data binning of 2×2 for the image mode and 1×2 in the long-slit mode was applied, respectively. The telescope was following the comet to recompense its proper velocity during the exposures. Consequently, direct images (hereinafter referred to as Ima) and long-slit spectra (hereinafter referred to as Sp) were obtained. The seeing (FWHM) was ~2″ (574 km) on April 14, 2020, and ~3″ (861 km) on April 16, 2020, respectively. The IDL codes created by the SAO RAS team were used for primary data reduction.

The observation approach used for the comet 2019Y4 observations, the detailed description of the image transformation, data reduction, and error estimation are the same as those explained in the papers by Afanasiev and Amirkhanyan (2012), Ivanova et al. (2015, 2016, 2017a, 2017b; 2019; 2020), Rosenbush et al., (2017, 2020), where also more details on the instrument can be found.

**2.2. Photometric Observations**

On April 14, 2020, the direct images of comet 2019Y4 were acquired with the narrow-band cometary filters: *BC* (λ4429/36 Å, from the Hale-Bopp set), *RC* (λ6835/83 Å, from the ESA filter set), and *SED500* (λ5019/246 Å). The *RC* filter was also used for observations on April 16, 2020. We observed the standard star G191B2B (Oke, 1990) for the photometric calibration of the images based on the SAO RAS spectral atmospheric transparency from the paper by Kartasheva and Chunakova (1978).

The reduction of the raw data consisted of removing the bias, the flat-field correction, and the cosmic ray removal. The flat-field image was taken at dawn for the non-uniform sensitivity of the CCD. The traces of cosmic rays were removed at the end of the data reduction through robust parameter estimation (Fujisawa, 2013). The level of the sky background was used in those parts of an image that were not covered with the coma and were deprived of faint stars, by the histogram of counts in the image. With the purpose to enhance the S/N ratio, we summed the comet images obtained during the night. The entire cycle of photometric observations during one night of observations took less than 4 hours. During this period, the fragments did not show any noticeable displacement relative to the bright component A (see Fig. 5). Therefore, summing the comet images separately in each filter is justified. The images of the comet were centered with the central outline of the relative intensity (isophotes) which was closest to the comet's maximum brightness. Coma morphology, gas production, dust colour, and normalized reflectivity obtained from the photometric data are in detail analyzed in Section 4.

**2.3. Spectroscopic Observations**

On April 14 and 16, 2020, we performed the long-slit spectroscopy of comet 2019Y4 using the VPHG1200@540 grism in the wavelength range of 3650-7300 Å. The spectral resolution was about 5.2 Å across the full range of wavelengths (a mean reciprocal dispersion was 0.89Å/px) and a spatial scale of 0.4″ per pixel along the slit. The spectrograph slit with the 6.8′×1.0″ dimensions was directed on the optocenter (i.e. the brightest point of the coma) and focused across the coma of the comet at position angles of 59.8° on April 14 and 83.4° on April 16 (see Fig. 1). The positional angle of the slit coincided with the positional angle of the comet's tail on April 16.

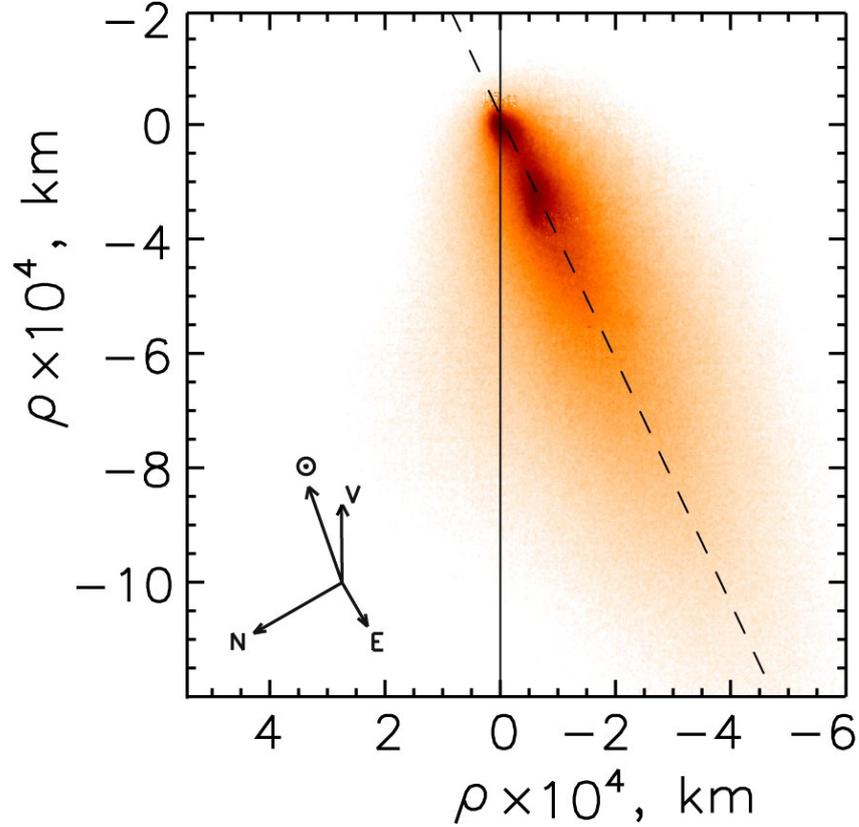

**Figure 1.** Orientation of spectrograph slit on April, 14 (the solid line) and April, 16 (the dashed line). The arrows indicate the directions to the Sun (☉), North (N), East (E), and the comet's velocity vector projected onto the sky plane (V).

We processed the obtained data using a standard procedure of the data reduction such as the bias and cosmic ray removing, and flat-fielding. We corrected every spectrum geometrically to compensate for the curvature of the spectroscopic image. We also performed the wavelength calibration, linearization, calibration of the flux, and we removed the spectrum of the sky. For flat-fielding, the spectrum of a built-in system of LEDs was used providing the almost uniform intensity distribution over a wide range of wavelengths (Afanaseiv et al., 2017). Since some strongest cometary emissions like CN 0-0 ($B^2\Sigma^+ - X^2\Sigma^+$) and $C_2$ 0-0 ($A^3\Pi_g - X^3\Pi_u$) cover the whole length of the slit, the night sky spectrum has been observed separately with similar exposure time and has been subtracted afterward. The He-Ne-Ar lamp was used for the wavelength calibration of the spectrum. For the cosmic ray removal and the S/N ratio improvement, we added and summed all the spectra with the robust averaging algorithm. We also observed the standard stars G19B2B and BD+75d325 from Oke (1990) for the absolute flux calibration of the spectrum of the comet. We took the spectral atmospheric transparency from the paper by Kartasheva and Chunakova (1978). Flux errors depend on the wavelength, the distance to the optocenter (or equivalently to the projected cometocentric distance ρ), and the date of observations. Errors were estimated as ~15% in the short-wave part of the spectrum, near 3% in the central, and up to 8% in the long-wave range of the spectrum. At the edges of the slit, where the comet signal is still present, the flux error was about 10% in the tail of the comet and up to 15% in the direction to the Sun. Table 1 shows the viewing geometry and the observation log of comet 2019Y4 for two time periods. The table presents the observation date and the UT range for each observation, the heliocentric distance (*r*) and geocentric distance (*Δ*), the phase angle (the Sun-Comet-Earth angle) (α), the filter or grism used, the total exposure time during the night ($T_{exp}$), and, finally, the observation mode.

**Table 1** Log of the observations of comet C/2019 Y4 (ATLAS) in 2020.

| Date, UT | *r* (au) | *Δ* (au) | α (deg) | Filter/grism | $T_{exp}$ (sec) | Mode |
|---|---|---|---|---|---|---|
| **April 14** | | | | | | |
| 17:29:58 – 18:29:47 | 1.215 | 0.999 | 52.8 | *SED500* | 3840 | Ima |
| 17:39:16 – 18:31:28 | 1.215 | 0.999 | 52.8 | *BC* | 4500 | Ima |
| 17:42:37 – 18:34:59 | 1.214 | 0.999 | 52.8 | *RC* | 4500 | Ima |
| 18:43:25 – 19:08:06 | 1.214 | 0.999 | 52.8 | *VPHG1200@540* | 7500 | Sp |
| **April 16** | | | | | | |
| 18:51:09 – 19:07:37 | 1.195 | 0.995 | 54.4 | *RC* | 6480 | Ima |
| 19:23:02 – 19:47:47 | 1.194 | 0.995 | 54.5 | *VPHG1200@540* | 4800 | Sp |

## 3. Analysis of the Observed Spectra

As we know, the observed spectrum of a comet is a combination of the emission spectrum of the coma gas and the spectrum of the sunlight scattered by dust particles. To separate the continuum signal from the gaseous emissions, we applied a method similar to that described in Ivanova et al. (2018). We used the Neckel and Labs (1984) high-resolution solar spectrum which was transformed to the resolution of our spectroscopic observations using convolution with the instrumental profile and normalization to the flux of the comet. We compared the convolved solar spectrum and the spectrum of the comet obtained on April 14, 2020 (Fig. 2a). Then we calculated a polynomial from the flux measured in both solar and comet spectra in the continuum windows: 3800 — 3830 ÅÅ, 4135 — 4165 ÅÅ, 4230 — 4260 ÅÅ, 4417 — 4483 ÅÅ, 5232 — 5288 ÅÅ, 5760 — 5820 ÅÅ, 6150 — 6200 ÅÅ, 6420 — 6460 ÅÅ, and 7099 — 7157 ÅÅ with the help of the IDL software robust_poly_fit.pro which made an outlier-resistant polynomial fit. The degree of the polynomial fit has been selected based on minimum absolute deviations. The polynomial (or redding) for both observation dates is shown in Fig. 2b. Multiplying the solar spectrum by this polynomial we obtained the continuum spectrum (the synthetic dust spectrum) which has been subtracted then from the observed spectrum. After subtracting, we have obtained a pure emission spectrum of comet 2019Y4. This result is displayed in Fig. 2c for both observation dates. It should be noted that the spectra of comet 2019Y4 shown in Fig. 2 were calculated with the rectangular diaphragm with a length of 10,000 km and a width equal to the slit size and centered at the maximum brightness of the spatial profile (which presumably contains the cometary nucleus).

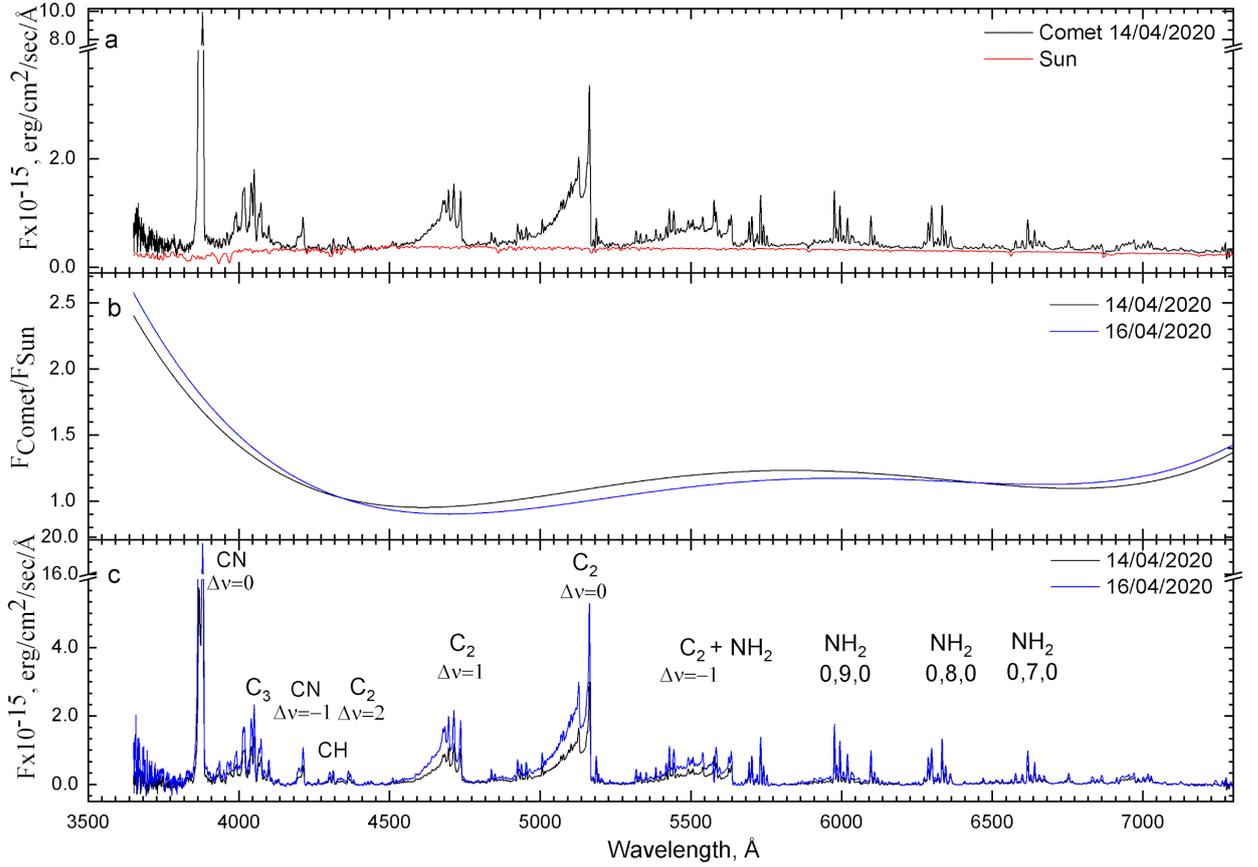

**Figure 2.** Reduction of the spectra of comet C/2019 Y4 (ATLAS): (a) the energy distribution in the comet's spectrum (the black line) obtained on April 14, 2020, with the rectangular diaphragm with a half-width of about 10,000 km and a length equal to the width of the slit and centered at the maximum brightness of the spatial profile and the shifted normalized spectrum of the Sun (the red line); (b) the normalized spectral dependence of the reflectivity of the dust (the black line — for observations of April 14, 2020, and the red line — for observations of April 16, 2020); (c) the emission component in the comet's spectrum (the black line — for observations of April 14, 2020, and the blue line — for observations of April 16, 2020).

Using this method of the synthetic dust spectrum calculated for every line of the slit image we have obtained the continuum signal for the whole slit. Given the value of the continuum signal for the whole slit, we can obtain the gas component. Fig. 3 (a,b) contains the contribution of the gas component as the ratio of the emission component to the total flux in each pixel of the whole slit for both observation dates.

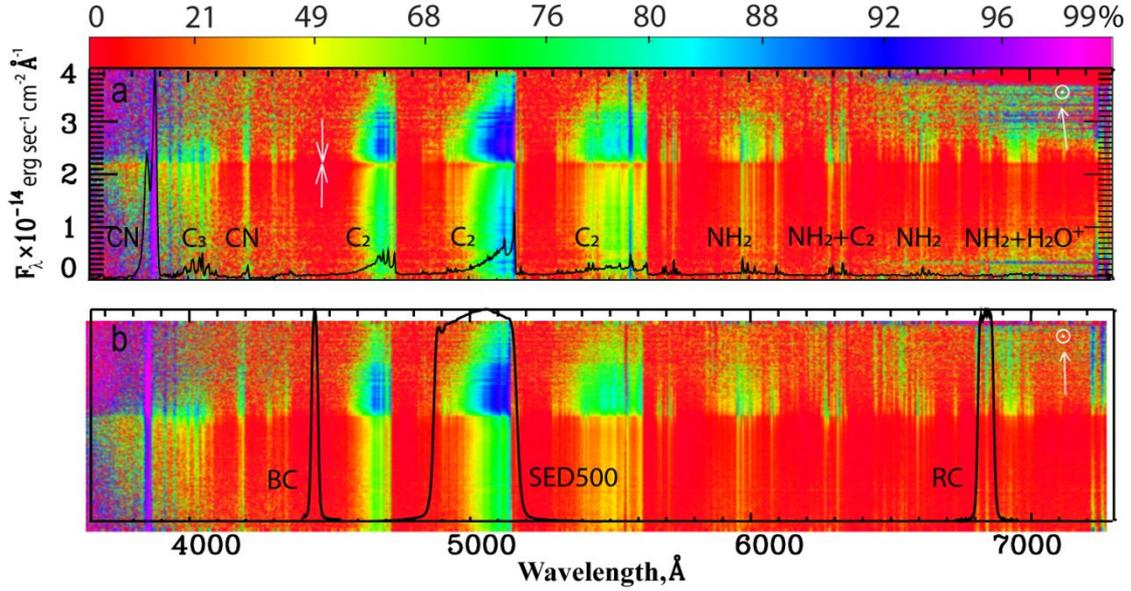

**Figure 3.** Spectroscopic observation of the comet C/2019 Y4 (ATLAS) on April 14 (a) and April 16 (b) 2020. The contribution of the gas component to the total flux (in percent) in every pixel of the image of the entire slit is colour-coded according to the colour bar on the top. The cometary spectrum (flux density in erg·sec$^{-1}$·cm$^{-2}$·Å) from April 14 is superimposed in panel (a) as a black line. Similarly, the normalized transmission curves of the *BC*, *RC*, and *SED500* filters are superimposed in panel (b) as the black line. The up-down white arrows on panel (a) point to the optocenter (presumably the nucleus or the brighter fragment of the comet). The white arrows on both panels indicate the projection of the direction to the Sun (☉) on the slit (see Fig.1).

The optocenter (presumably the nucleus or the brighter fragment of the comet) with a reduced gas contribution is manifested in Fig. 3. The increased gas contribution areas coincide with the emission spectral ranges of the main cometary gas emissions (see the spectrum of comet 2019Y4 plotted in the slit image). But it should be noted that the distribution of the contribution of the gas component on the whole slit is asymmetric concerning solar-anti-solar direction. Fig. 3 also shows that the distribution of the contribution of the gas component on the whole slit depends on the orientation of the spectrograph slit. This also indicates the non-isotropic and heterogeneous outflow of the matter. Since the contribution of the gas component was calculated as its ratio to the total flux, which is the sum of the gas and dust fluxes, the asymmetry is also inherent in the dust component.

The transmission curves of the *BC* and *RC* filters in the slit image are shown in Fig. 3b. As one can see, the *RC* filter is strongly contaminated with gas emissions in the Sun direction. Fig. 4a shows the value of the contribution of the gas component depending on the distance to the nucleus of the comet. The contribution of the gas component was estimated as the ratio of the convolution of the emission component with the transmission curve of the corresponding filter at a certain distance from the nucleus to the convolution of the total flux with the transmission curve of the corresponding filter at the same distance from the nucleus. As one can see, for the *BC* filter in the anti-solar direction, the gas contribution is comparable to the error, while for the *RC* filter fluctuates within 7 — 8%. But in the Sun direction, the situation with the *RC* filter is getting worse.

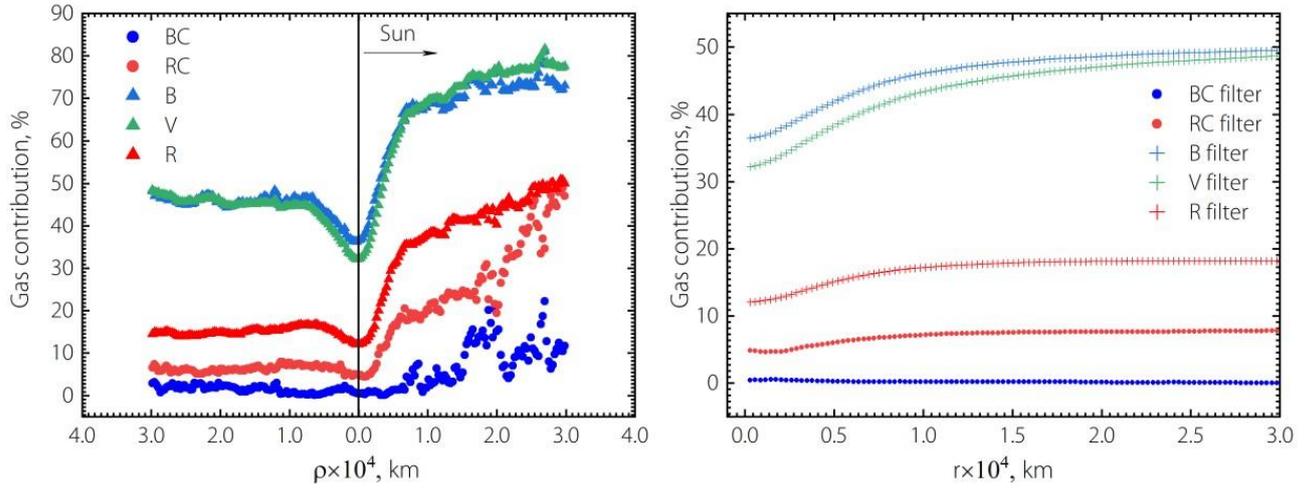

**Figure 4.** Contribution of the gas component to the total flux through the *BC* (the blue dots) and *RC* (the red dots) filters depending (a) on the cometocentric distance and (b) on the aperture radius in comet C/2019 Y4 (ATLAS) from the observation on April 14, 2020. The contributions of the gas component to the total flux through the commonly used broad-band Johnson-Cousins filters are shown for comparison.

To determine the degree of the emission contamination of the filters more accurately, we calculated the contribution of the gas component depending on the aperture size (Fig. 4b). As we can see, the gas contribution is insignificant for the *BC* filter, and for the *RC* filter, it spans from 5% to 8%, depending on the size of the aperture. The splitting of the comet demonstrates an increase in the presence of the gas component of the cometary coma in the RC filter, primarily due to the presence of the emissions $NH_2$ and $H_2O^+$, as can be seen from the comparison of panels a and b in Fig. 3.

Since the broad-band filters of the Johnson-Cousins system are often used to determine the proxy of the dust abundance within the coma (*Afρ*), we also calculated the contribution of the gas component for these filters. For the *R* filter, this contribution increases from 12% to 18% with the increasing aperture radius. While for the *B* and *V* filters, it increases up to almost 50% for large apertures. The nucleus splitting makes this comet quite gaseous.

The Haser model, which is used to determine the rate of gas production in comets, and *Afρ* are based on an assumption of the continuous and isotropic outflow of gas and dust from the nucleus (i.e., the coma is spherically symmetric). However, the Haser model, despite the lack of a physical basis, empirically fits the observed gas distribution (A'Hearn et al., 1982). Accordingly, we use this model to evaluate the gas production in comet 2019Y4. A'Hearn with colleagues (A'Hearn et al., 1984) introduced the parameter *Afρ* as a parameter for the proxy of the dust abundance within the coma. In Ivanova et al. (2018), it was noted that recovering the dust production from the observed *Afρ* parameter is a doubtful task. The success of this method depends on the dynamical and optical characteristics of the dust grains (e.g., the effective density and cross-section as well as the scattering coefficient and phase function). To bypass these problems, the method of determination of the dust amount in the coma within the selected aperture from its brightness is proposed in Zubko et al., (2020). We did not determine the dust production of the comet 2019Y4 using the *Afρ* parameter.

4. **Analysis of the Photometric Data**

Observations of the comet 2019Y4 taken on April 14 and 16, 2020 were performed 47 and 45 days before its perihelion passage on May 31, 2020, respectively. Three narrow-band images of comet 2019Y4 shown in Fig. 5 were taken with the cometary continuum *BC* and *RC* filters and in the *SED500* filter centered on the (0–0) transition of the $A^3\Pi_g$–$X^3\Pi_u$ electronic band system of the $C_2$ molecule.

In the *BC* and *RC* filters, the comet displays an elongated coma in the solar direction and four round condensations (A, B, C, and D) perhaps surrounded by weak small comae. The cometary coma is compact in the solar direction; this suggests that the dust found in comet 2019Y4 is located predominantly close to the nucleus (a parent fragment). The cometary coma in the *RC* filter is more extended than that in the *BC* filter. Fragments C and D in the *BC* filter exhibit a much larger size and are shifted relative to the solar-tail axis.

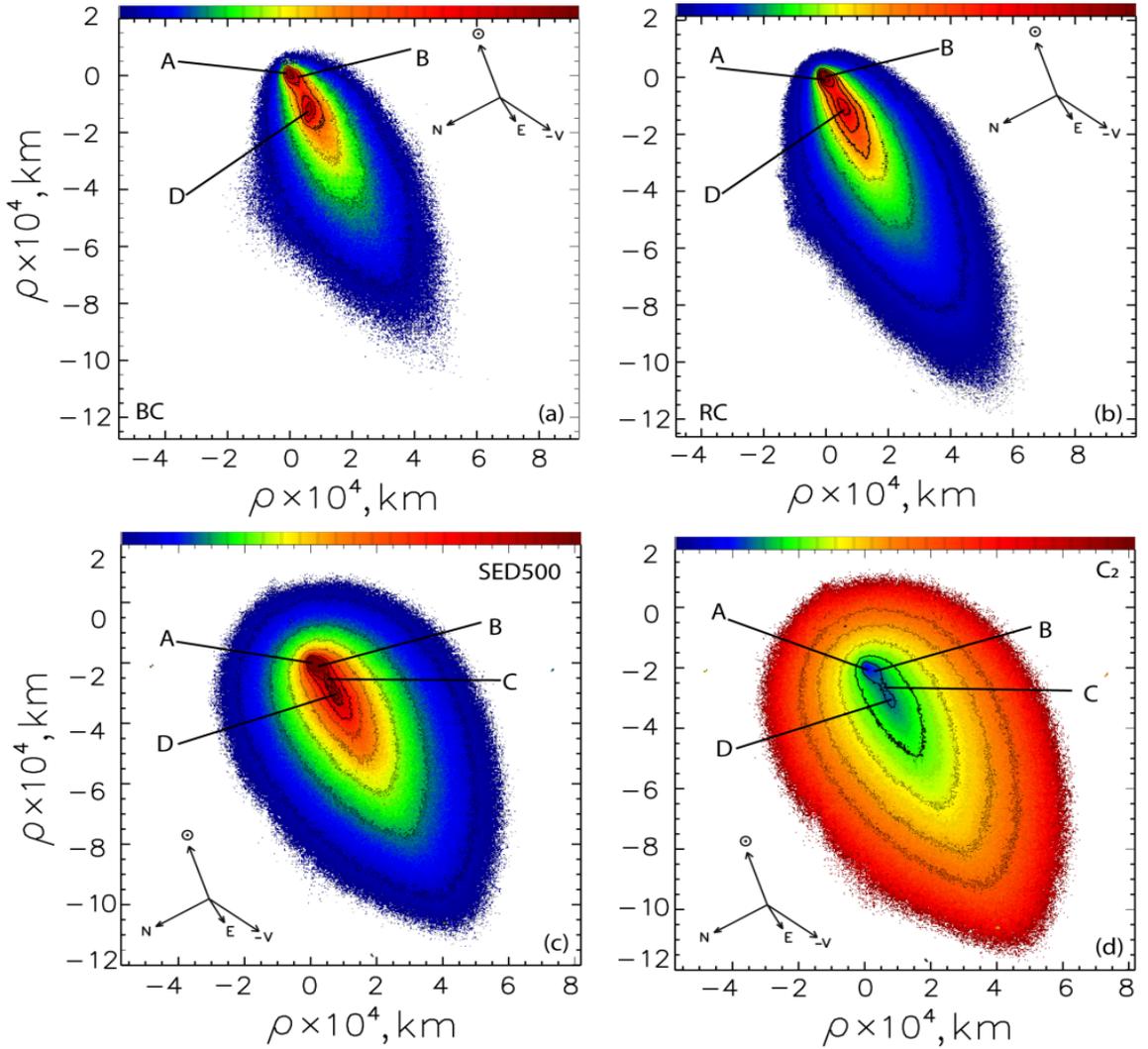

**Figure 5.** Images of comet C/2019 Y4 (ATLAS) acquired in the narrow-band filters: the blue continuum *BC* (a), the red continuum *RC* (b) filters, and the filter focused at the $C_2$ band system (c) on April 14, 2020. Diagram (d) shows the pure $C_2$ intensity map after the dust component subtraction. North, East, sunward, and the negative velocity vector directions are marked.

The *SED500* and pure $C_2$ emission images of comet 2019Y4 are symmetric relative to the parent component and a bit prolonged in the tail. To obtain the pure emission image of the comet, the continuum was removed. To perform the subtraction correctly, we have considered the different continuum levels from the spectroscopic data obtained and the fact that the filter light transmissions are different. In Fig. 5(d), the pure $C_2$ intensity map after the dust component subtraction is given. Comparing diagrams (c) and (d), we can conclude that the dust does not contribute considerably to the surface brightness of the $C_2$ image.

### 4.1 Coma Morphology with Narrow-band Filters

To distinguish low-contrast structures found in the images, we performed different methods of digital processing: the rotational gradient technique given in the paper by Larson and Sekanina, 1984, and the division by azimuthal average filtering method described in the paper by Samarasinha and Larson, 2014. To eliminate any false features while interpreting the images, we applied every digital filter with similar reduction parameters to all separate exposures as well as to the co-added image according to the method of Manzini et al (2007). The present method has been already successfully used for the separation of structures found in some comets (Ivanova et al., 2017b, 2019; 2020; Rosenbush et al., 2017; 2020).

Fig. 6 gives the intensity images of comet 2019Y4 processed in two different digital techniques. Panels (a), (b), and (c) represent the results of the rotational gradient technique (Larson and Sekanina, 1984; Samarasinha and Larson, 2014) applied to the observations from April 14, 2020, in SED500, BC, and RC filters. Panel d) is the result of the same technique applied on the intensity observation from April 16, 2020. Panels (A), (B), and (C) show the intensity image processed by the method of the division by azimuthal average (Samarasinha and Larson, 2014). We used bright component A (see Fig. 5) as an optocenter for the application of digital filters.

The comet displayed an extended coma with highly condensed material in the near-nucleus area and a tail in the anti-solar direction with a length of approx. $1\times10^5$ km measured in the projected nucleocentric direction. The comparison between the *BC* and *RC* images reveals that the comet morphology is the same in both filters, but in the *BC* filter, the coma that is produced by the dust looks fainter than that for the *RC* image. Fig. 6 illustrates the fact that all the images show similar morphology of the coma, which means that a notable tail is observed in the anti-sunward region directly focused on the sunward vector in all the photometric bands.

Both filters BC and RC do not transmit the emission radiation of the gas and are therefore used for the analysis of the dust coma morphology. Again, we do not exclude that the splitting of the comet can increase the gas component of the cometary comae, and consequently, the *BC* and *RC* images can be contaminated by the gas emissions (see Figs. 3 and 4).

In the processed images (the right-hand panel in Fig.6), apart from the dust tail, the well-defined morphological features (four components of the splitting comet) are revealed. These components are seen in all our images at approximately the same position angles during the whole observation period. Small changes in the dust coma morphology detected in the filters BC and RC are probably due to the intrinsic variability of the dust cometary coma.

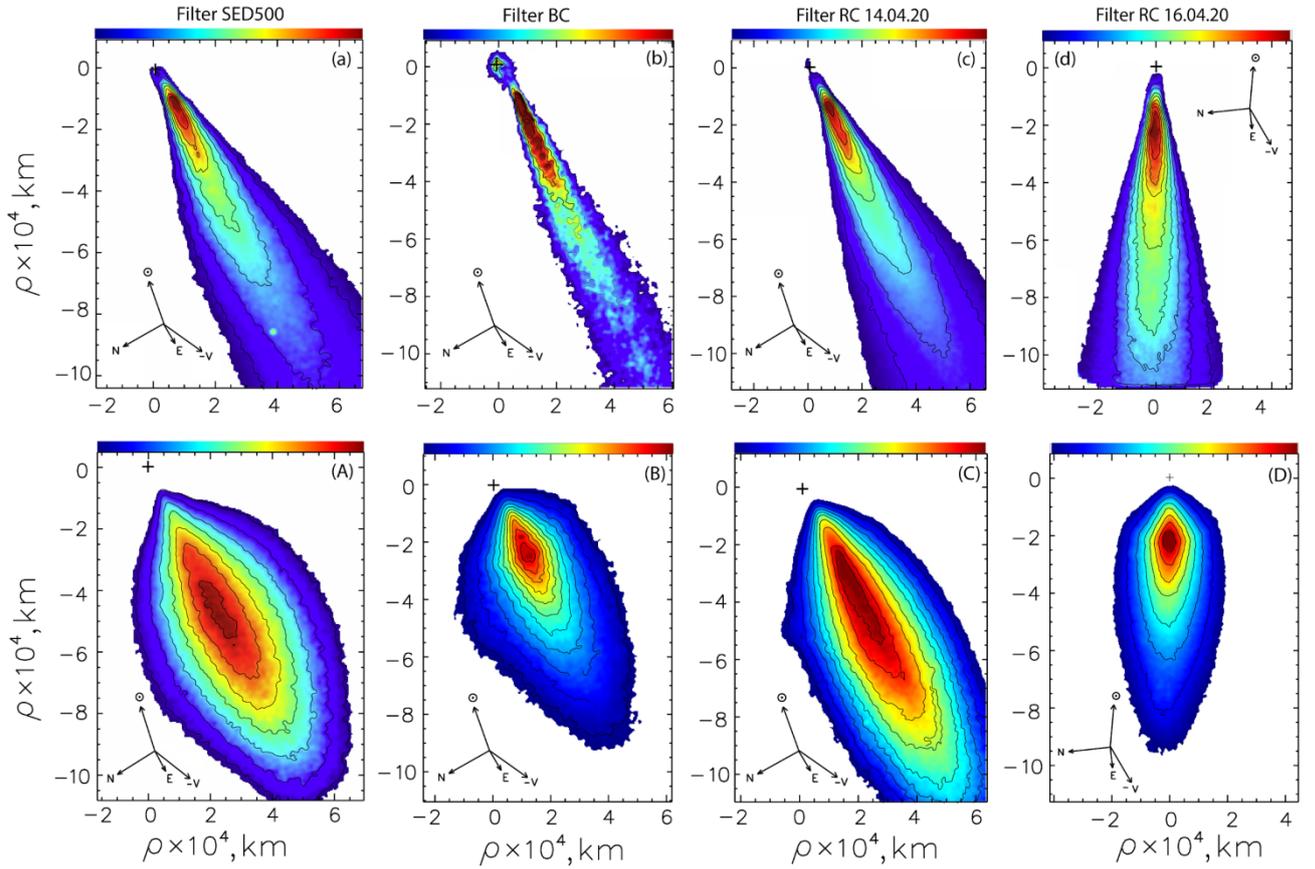

**Figure 6.** Top panels: Intensity images of comet C/2019 Y4 (ATLAS) in *SED500, BC*, and *RC* filters (a, b, and c) from April 14, 2020, and (d) image in *RC* filter from April 16, 2020, obtained by the rotational gradient method (Larson and Sekanina, 1984; Samarasinha and Larson, 2014). Bottom panels: Intensity images of the comet in *SED500, BC*, and *RC* filters (A, B, and C) from April 14, 2020, and (D) image in *RC* filter from April 16, 2020, obtained by the method of azimuthal average division. North, East, sunward, and negative velocity vectors are indicated in each frame.

We detected only four major fragments (A, B, C, and D according to the Minor Planet Center identification), each probably consisting of 1 to 2 brighter components. Fainter fragments, which were found later with the Hubble Space telescope, were not detected on our images. Hui and Ye (2020), who observed the comet before our observation from January 19 to April 5, 2020, noted that the central optocenter of the comet had been significantly condensed until April 5, 2020, and afterward, it became even more diffuse and elongated. The morphological change proves that the nucleus of the comet has split into multiple fragments (Ye & Zhang, 2020). Steele et al. (2020) and Lin et al. (2020) affirm this aspect.

### 4.2 Rates of Gas Production

For the calculation of the production rate of gas molecules, we used the model defined by Haser (1957). The CN violet 0–0 and 0-1 ($B^2\Sigma^+ - X^2\Sigma^+$) bands, the $C_3$ ($A^1\Pi_u - X^1\Sigma_g^+$) group, the $C_2$ 0-0 ($A^3\Pi_g - X^3\Pi_u$) band, and the NH2 ($\tilde{A}^2A_1 - \tilde{X}^2B_1$) band can be recognized in the emission spectrum in Fig. 2c. We took the values of the fluorescence efficiency (the g-factor), the parent and daughter scale lengths ($l_p$ and $l_d$, respectively), and the power-low index $n$ dependent on the heliocentric distance ($r^{-n}$) from the paper by Langland–Shula and Smith (2011). The g-factor for the CN(0–0) band and the heliocentric velocity of the comet are dependent on the heliocentric distance. Tatum (1984) gave the fluorescence efficiency values for the CN molecule considering the Swings resonance-fluorescence effect. Schleicher (2010) published tables of the g-factor values for different heliocentric distances and velocities, and we used them in our calculations. In Table 2, we present the parameters we used for the Haser model.

Table 2. Model parameters that we used to determine the gas production rates

| Molecule | g-factor (watts mol$^{-1}$) | $l_p$ ($10^4$ km) | $l_d$, ($10^4$ km) | Power-law index $r^{-n}$ | Lifetime (sec) |
|---|---|---|---|---|---|
| CN(0–0) | $4.4 \times 10^{-20}$ | 2.19 | 30.00 | 2 | 487156.0 |
| C$_2$($\Delta v$=0) | $4.5 \times 10^{-20}$ | 1.60 | 11.00 | 2 | 178624.0 |
| C$_3$ | $1.0 \times 10^{-19}$ | 0.10 | 6.00 | 2 | 97431.2 |
| NH$_2$ | $2.29 \times 10^{-21}$ | 0.41 | 6.20 | 1.55 | 100679.0 |

The Haser model is based on the assumption of the continuous and isotropic outflow of gas and dust from the nucleus, and that the gas emission is centered in the nucleus. Fig. 5 shows that we are not dealing with a single nucleus, but with a chain of fragments (i.e., the coma is not spherically symmetric). Fig. 7 shows the spatial profiles of the main cometary emissions, for which we calculated the value of productivity according to the Haser model. Indeed, we can see from Fig. 7 that the process of splitting of the nucleus of comet 2019Y4 changes the form of the spatial profile of the comet emissions. There can be identified small secondary maxima, which correspond to the fragments of the splitting nucleus. For our calculations of the gas production rate, we chose an aperture (marked in Fig. 7 with dashed lines), which cuts out only the central nucleus and the near nucleus region (about 10,000 km). In this area, the influence of secondary fragments is minimal (or absent).

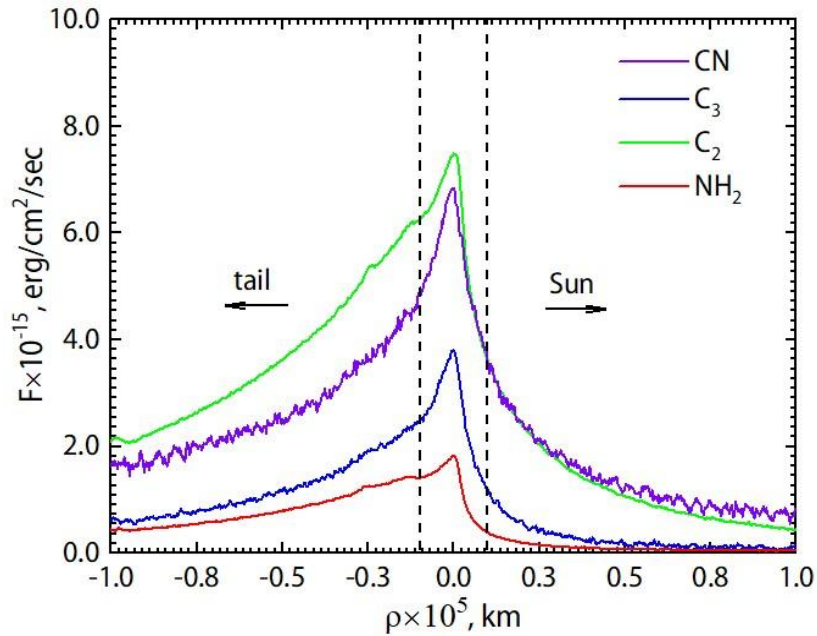

**Figure 7.** Spatial profiles of main comet emissions calculated using transmission profiles for the HB filters CN, $C_3$, $C_2$, and $NH_2$ from the spectrum of comet C/2019 Y4 (ATLAS) for observations of April 16, 2020. Arrows indicate the direction to the Sun and tail. The dashed vertical lines indicate the boundaries of the diaphragm (about 10,000 km), which was used for the calculation of dust productivity according to the Hazer model.

The Haser formula is defined for a circular aperture, and it has to be modified when used in the case of a rectangular aperture of the spectrograph slit. The formula was multiplied by a factor to compensate for the loss of signal due to the use of rectangular aperture concerning the use of circular one. With the use of parameters listed in Table 2, we computed the production rates of CN, $C_3$, $C_2$, and $NH_2$ in the comet 2019Y4. Based on the composition defined through the ratio of the production rates of different species, A'Hearn et al. (1995) divided comets into two classes: depleted and typical. The depleted comets are characterized by $\log[Q_{C_2}/Q_{CN}]<–0.18$ whereas the mean value of this ratio for typical comets is 0.06. As we can see from the $\log[Q_{C_2}/Q_{CN}]$ values listed in Table 3, comet 2019Y4 can be classified as a typical comet.
Langland-Shula and Smith (2011) considered the value of the logarithmic ratios of carbon-chain to CN production rate for comets of different dynamical categories. They showed that some of the comets of the Jupiter-family and the Long-period comets are depleted in $C_2$ relative to CN, and that the comets of the Jupiter-family are more depleted in $C_2$ than the Long-period comets. Langland-Shula and Smith (2011) derived the mean values of $\log[Q_{C_2}/Q_{CN}]$ for different dynamic types: –0.7 ± 0.2 for Jupiter-family, 0.02 ± 0.01 for Halley-type family, –0.1 ± 0.1 for Long-period comets, and 0.2 ± 0.2 for Dynamically new comets. Our results show that comet 2019Y4 belongs to the category of Long-period comets in the dynamic classification of Langland-Shula and Smith (2011).

**Table 3.** Gas production rates for comet C/2019 Y4 (ATLAS), obtained for spectral observations

| UT Date, 2020 | $r_h$, [au] | $\log \rho$, [km] | $\log Q_{CN}$ [mol s$^{-1}$] | $\log Q_{XX}/Q_{CN}$ | | | $\log F/F_{CN}$ | |
|---|---|---|---|---|---|---|---|---|
| | | | | $C_3$ | $C_2$ | $NH_2$ | $CO^+$ | $H_2O^+$ |
| Apr., 14.74 | 1.21 | 4.003 | <25.02 | –1.84±0.1 | –0.1±0.1 | –0.6±0.1 | –1.8 | –1.17 |
| Apr., 16.77 | 1.17 | 4.006 | <25.05 | –1.86±0.1 | –0.15±0.1 | –0.57±0.1 | –1.89 | –1.12 |

The $C_2$ molecules dramatically contaminate the cometary spectra in the visible range and the effect is most evident in the broad-band *V* filter of the Johnson photometric system. It is possible to compute the column density of the $C_2$ molecules (Langland–Shula and Smith, 2011) from spectroscopic observations using the Haser model. We obtained the $N_{C_2} = 2.18972 \times 10^{28}$ molecules for an area of the aperture of the spectrograph slit with a center on the optocenter (presumably the nucleus or the brighter fragment of the comet) corresponding to a diameter of about 10,000 km of the circular diaphragm. The absolute magnitude of a single $C_2$ molecule ($\Delta v = 0$) within the *V* filter passband and scaled to heliocentric and geocentric distances of comet 2019Y4 for April 14, 2020, is 85.783 mag. We calculated the cumulative magnitude of the $C_2$ molecules ($\Delta v = 0$) for comet 2019Y4 within an aperture diameter of about 10000 km as 15.3$^m$. At the same time, the brightness of the comet on April 15.47, 2020 in the *V* filter measured with an aperture radius of 6000 km was 15.83$^m$ (Zubko et al. 2020). Considering that here the aperture diameter is slightly larger, which means that for a smaller aperture, the comet's brightness in the *V* filter from the photometric observations will be even lower. The result can be applied to other cometary molecules. It means that the obtained values of the production rates for the main cometary molecules in comet 2019Y4, from spectroscopic observations, can be regarded as the maximum values.

On April 14, the image of comet 2019Y4 in the *SED500* filter which corresponded to the emission range of the $C_2$ molecule ($\Delta v = 0$) was also obtained. The same parameters of the Haser model for an aperture with a diameter of about 10,000 km were used. The value of the productivity of the $C_2$ molecule from the photometric observations $\log(Q_{C2}) = 23.3$. The value confirms the conclusion about the upper limit of the defined values of the productivity for the main cometary molecules. The slit covered a region on the comet's head with a higher integral brightness which ultimately led to an overestimation of the gas productivity.

The correlation between the findings on the composition of the gaseous coma in 2019Y4 with that emerging from the study of its dusty coma (Zubko et al. 2020) is worth noting. The CN and $C_2$ molecules in a cometary coma could evaporate from the submicron and/or micron-sized organic particles in the cometary jets (e.g., A'Hearn et al. 1986; Cosmovici et al. 1988). Such particles were found in the comets *in situ*; in the literature, they are referred to as *CHON* particles (e.g., Fomenkova et al. 1992). On the other hand, the ground-based imaging polarimetry of comets suggests that their jets do not reveal the phenomenon of negative polarization at small phase angles (e.g., Hadamcik and Levasseur-Regourd 2003; Hines et al. 2014). The absence of the negative polarization is an indicator of the predomination of highly absorbing particles, whose refractive index corresponds to the organic materials (Zubko et al. 2012) or can result in the scattering by small dust particles (Rayleigh scattering). This result may explain the high abundance of the $C_2$ and CN molecules detected in progenitors (Fig. 3).

## 4.3 Colour and Normalized Reflectivity

We used the images obtained on April 14, 2020, in the narrow-band *BC* and *RC* filters of the cometary continuum to calculate the dust colour of the comet 2019Y4. We carefully centered the comet images obtained in both bands and calibrated the images for the flux at the same position by the central brightness peak of the coma determined from the isophotes. Using the images obtained in this way, a colour map was constructed. The map can be used for the analysis of the colour variations in the coma. To finish the colour map, we also transformed each pixel of the summed images into the apparent magnitude. The measurement error of the mean magnitude equals $0.1^m$ g. Fig. 8a presents the (*BC–RC*) colour map of comet 2019Y4. The right panels of Fig. 8 show the slices cut across the comet colour map along the tail (b) and in the direction perpendicular to the tail (c). As the distance from the nucleus increases, the colour profiles show the asymmetry in the colour gradient. The coma colour shows a steep weakening from approximately $1.4^m$ (although, not symmetrically to the nucleus) to approximately $0.0^m$ at a distance of ~20 000 km in the Sun direction and the smooth rise up to approximately $1.5^m$ at a distance of ~100 000 km in the tail direction. A significant difference in the behavior of the colour in Fig. 8a is observed. Most probably, the difference is due to the asymmetry of the contribution of the gas component to the total flux along the sun-tail direction (see Fig. 2). The moderately red intrinsic colour of the dust near the nucleus turns into immensely blue (a colour-index of approximately $~1.0^m$) further from the nucleus. Solar colour-index *BC–RC* =1.276 (Farnham et al., 2000) was used.

As can be seen in Figs. 8b and 4a, both colour and contribution of the gas component correlate with the Sun direction. For example, the colour rapidly changes from red to blue (from $1.37^m$ to $0^m$ at a distance of 70,000 km from the nucleus), and the contribution of gas increases rapidly with increasing distance from the nucleus. At the same time, the contribution of the gas component is insignificant in the anti-solar direction (along the tail) and practically does not change with cometocentric distance. The colour of the coma has a reddish trend in the anti-solar direction (redding of the coma colour from $1.37^m$ to $1.47^m$).

The above properties suggest some evolution of the scattering particles, presumably, as a consequence of some changes. It should be noted that a similar result (reddening in the anti-solar direction) was obtained by Li et al (2013) for the comet C/2012 S1 (ISON). In Zubko et al (2015), the effect of coma reddening in anti-solar direction is explained as the effect of the radiation pressure that decreases index n in the power-law size distribution $r^{-n}$ of cometary grains. In the direction perpendicular to the Sun-tail (Fig.8c), we can see a blueing trend of the coma colour. From Fig. 5 we can see that in the direction perpendicular to the sun-tail direction there is fragment B, which is slightly shifted towards the tail. This fragment exhibits quite strong activity, which can lead to the asymmetry that we see in Fig. 8c.

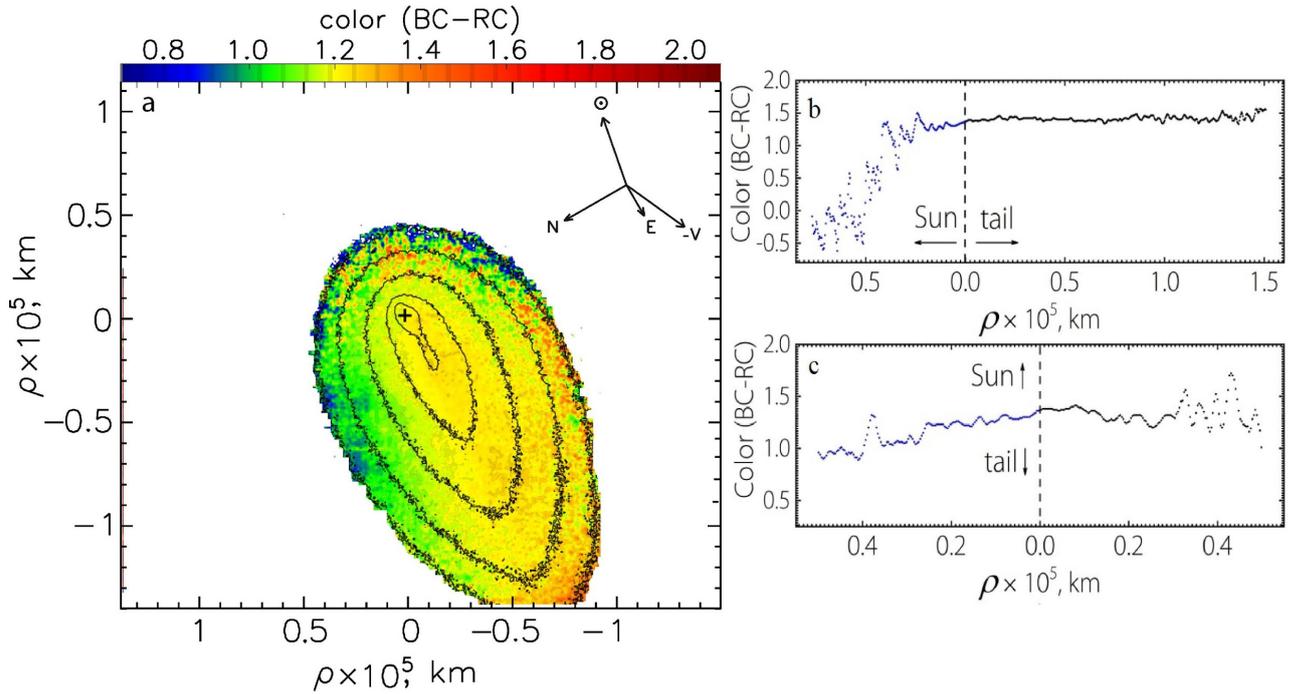

**Figure 8.** The (*BC–RC*) colour map of comet 2019Y4 (a) and the slices cut across the comet colour map along the tail (b), and in the perpendicular direction to the tail (c) as a function of the projected cometocentric distance, $\rho$. Results obtained from the photometric measurements on April 14, 2020.

It should be noted that the blueing trend of the coma colour from the nucleus to the coma periphery (Fig. 8b and 8c) affects the determination of the colour of the comet's coma in the case of aperture photometric measurements. The coma appears bluer with the increasing size of the aperture. There are published only a few estimations of the colour of the comet 2019Y4. Ye, Q., & Hui, MT (2020) showed that in March 2019, the comet had a blueing trend from a reddish colour $(g − r) \approx 0.6^m$, in comparison to the Sun's $(g − r) = + 0.46^m \pm 0.04^m$ (Willmer, 2018) since January 2020, reached a dip at an epoch of ~60 d preperihelion, when the comet appeared even bluer than the Sun $(g−r) \approx 0.2^m$ and began to be reddening afterward. They concluded that the comet began to split in March 2019. Our narrow-band photometric observations give a more neutral colour (the colour index: $BC–RC \approx 0.1^m$). Kochergin et al. (2020) and Zubko et al. (2020) carried out broad-band photometric and polarimetric observations of the comet during similar epochs. A blue colour (Kochergin et al., 2020) (the colour index: –0.2 $^m$ in mid-April and –1.0 $^m$ at the end of April) and very high positive polarization (Zubko et al., 2020) of the comet was observed. Observed colour differences are most probably caused by the use of filters of different photometric systems, which, however, may be useful for determining physical characteristics of the dust (Lukyanyk et al., 2020). Nevertheless, all these results indicate the occurrence of non-stationary processes associated with the nucleus splitting.

The reflectivity S'(λ) variations along the dispersion expressed as the comet spectrum $F_{com}(\lambda)$ divided by the scaled spectrum of the Sun $F_{sun}(\lambda)$: $S'(\lambda) = F_{com}(\lambda)/F_{sun}(\lambda)$ were studied. The polynomial from Fig. 2b was used to determine the dust reflectivity. The result indicates that the dust reflectivity increases with the growing wavelength. It may be quantitatively shown as the normalized reflectivity gradient using the expression:

$$S'(\lambda_1, \lambda_2) = \left(\frac{2000}{\lambda_2 - \lambda_1}\right) \frac{(10^{0.4\Delta m} - 1)}{(10^{0.4\Delta m} + 1)},$$

where S'(λ$_2$) and S'(λ$_1$) comply with the measurements performed at the wavelengths $\lambda_1$ and $\lambda_2$ (in Å) with $\lambda_2 > \lambda_1$. Gradient $S'(\lambda_1, \lambda_2)$ is expressed in percent per 1000 Å. Values of the calculated reflectivity gradients within the wavelength region of 4429 — 6835 Å (BC-RC) are listed in Table 4. For a comparison, Storrs et al. (1992) derived the average value of about 22% per 1000 Å (with the minimum of 15% per 1000 Å and the maximum of 37% per 1000 Å) within the wavelength region of 4400 — 5600 Å for 18 ecliptic comets. It should be noted that spectral gradients from 3.6% per 1000 Å to 21.8% per 1000 Å were obtained for different components of the splitting comet 73P/Schwassmann-Wachmann 3 (Bertini et al., 2009).

**Table 4.** The normalized spectral gradient in Comet C/2019 Y4 (ATLAS), obtained from spectral observations

| UT Date, 2020 | $r_h$, [au] | log ρ, [km] | S', [% per 1000 Å] BC — RC |
|---|---|---|---|
| Apr., 14.74 | 1.21 | 4.003 | 17.3/15.5* |
| Apr., 16.77 | 1.17 | 4.006 | 17.7 |

*- from photometry

## 5. Mutual Evolution of comet C/2019 Y4 (ATLAS) and C/1844 Y1 (Great comet)

After the discovery of comet 2019Y4 by the Asteroid Terrestrial-Impact Last Alert System (ATLAS) at Mauna Loa, Hawaii, the amateur astronomer Maik Meyer reported that the 2019Y4 orbit resembles the C/1844 Y1 (Great comet) (hereafter, 1844Y1) orbit (https://groups.io/g/comets-ml). Similar orbits of comets could point at their possible common origin. The dynamical and geometrical similarity of the orbital elements of comets indicates a common origin of the corresponding nuclei despite that the perihelion passage times differ considerably. The comets can have a common origin even though it is not possible to identify the time and location of the splitting using the simple numerical backward integration of their orbits (Marsden and Sekanina, 1971). The orbital elements are similar for several possible pairs of split comets: C/1988 F1 (Levy) and C/1988

J1 (Shoemaker-Holt) (Bardwell, 1988); C/1988 A1 (Liller) and C/1996 Q1 (Tabur) (Jahn, 1996); C/2002 C1 (Ikeya-Zhang) and C/1661 C1 (Green, 2002); C/2002 A1 (LINEAR) and C/2002 A2 (LINEAR) (Sekanina et al., 2003).

We performed the numerical integration of the 1844Y1 and 2019Y4 orbits in the past and in the future in order to study the dynamical relations between the comets. For our purpose, the orbital elements for both comets were taken from the JPL Small-Body Database Browser (Giorgini et al., 1996). The heliocentric ecliptic orbital elements of 1844Y1 relating to the equinox J2000.0 are: $q$ = 0.2505 au, $a$ = 358.9355 au, $e$ = 0.9993, $\Omega$ = 120.5910°, $\omega$ = 177.5055°, $i$ = 45.5651° for epoch 2394620.5 and the heliocentric ecliptic orbital elements of C/2019 Y4 are: $q$ = 0.2528 au, $a$ = 330.6077 au, $e$ = 0.9992, $\Omega$ = 120.5721°, $\omega$ = 177.4084, $i$ = 45.3839° for epoch 2458915.5.

The numerical integration of the comet orbits was performed with the RA15 integrator (Everhart, 1985) within the MERCURY package (Chambers, 1999). The gravitational perturbations of eight planets (from Mercury to Neptune) were considered; however, non-gravitational effects were ignored in the integration.

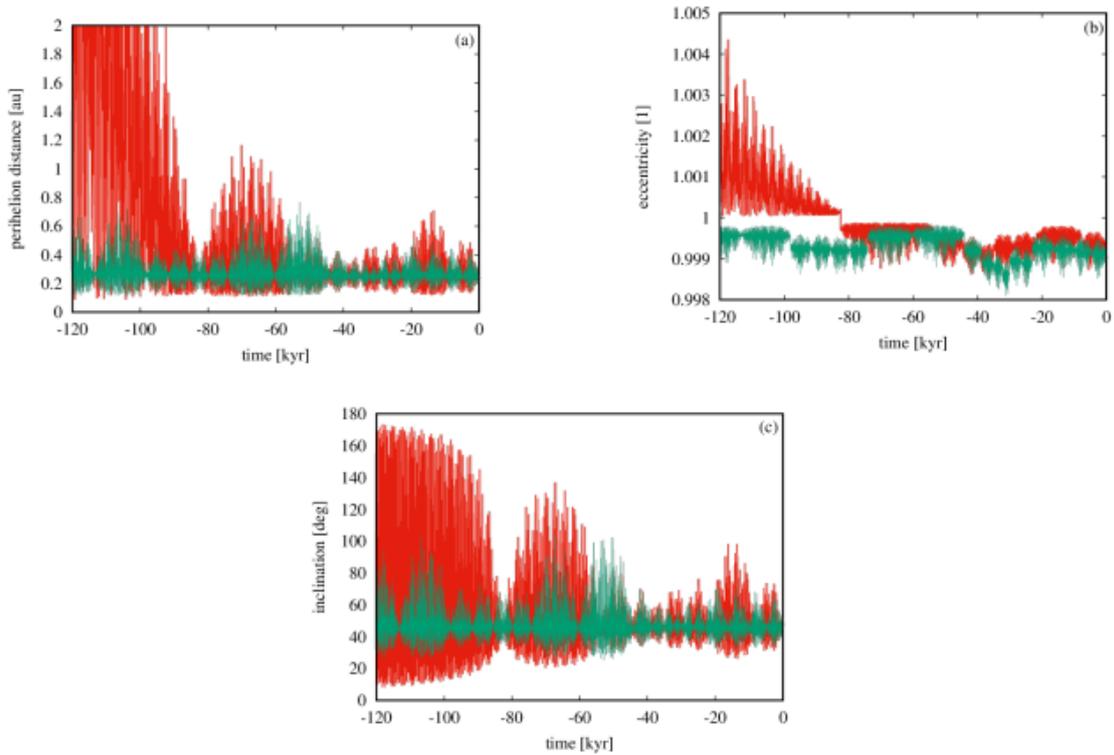

**Figure 9.** Evolution of the perihelion distance (a), the eccentricity (b), the inclination to the ecliptic plane (c) of the orbit of C/1844 Y1 (Great comet) (the red curve), and the orbit of C/2019 Y4 (ATLAS) (the green curve) for up to 120 kyr in the past.

The evolution of the orbits of comets 1844Y1 and 2019Y4 backward in time for 120 kyr are shown in Fig. 9. We present a set of the orbital elements, the perihelion distance (Fig. 9a), the eccentricity (Fig. 9b), the inclination to the ecliptic plane (Fig. 9c). The red curve represents comet 1844Y1 and the green curve is for comet 2019Y4. According to the evolution of the eccentricity for comet 1844Y1, we can assume that the orbit of the comet had been hyperbolic and, under the influence of the gravitational forces, it had changed to the elliptic ~82 kyr ago. At that time, the comet had a close approach to Mercury at a distance of 0.038 au which could have caused the comet's orbit to change from hyperbolic to elliptic.

The later evolution of the orbital elements for both comets is relatively similar. The evolution of the orbit of comet 2019Y4 is less turbulent than that of comet 1844Y1. There are no significant changes as in the case of comet 1844 Y1. It keeps a certain average value of the orbital elements during the investigating period. The identity of the orbits of 1844Y1 and 2019Y4 was investigated by Hui & Ye (2020). Using numerical integration of the comets, they found that the split event in the comet pair of 1844Y1 and 2019Y4 happened around the progenitor's previous perihelion passage (~5 kyr in the past). 1844Y1 was formed. Hui & Ye (2020) integrated nominal orbits and orbits of 1,000 clones for either of the comets backward ~7 kyr. Indeed, the orbits of both comets were very similar according to our integration at this time, but comet 1844Y4 was in the hyperbolic orbit more than ~82 kyr ago as was mentioned above.

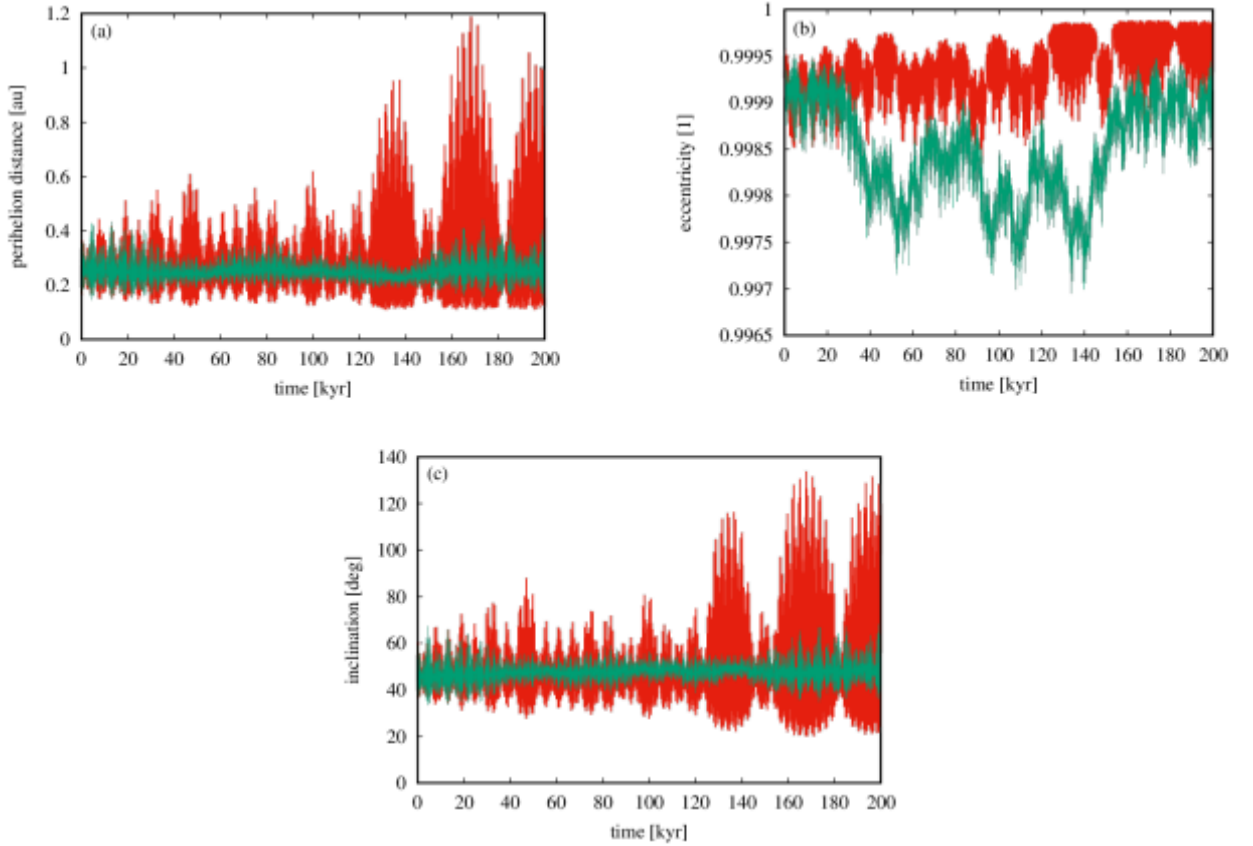

**Figure 10.** Evolution of the perihelion distance (a), the eccentricity (b), the inclination to the ecliptic plane (c) of the orbit of C/1844 Y1 (Great comet) (red line), and the orbit of C/2019 Y4 (ATLAS) (green line) for up to 200 kyr in the future.

The evolution of the orbital elements for both comets will be quite similar in the next tens of thousands of years in the future (Fig. 10a). The evolution of the perihelion distance and the inclination for both comets are similar for a longer time (approximately 125 kyr). Significant differences will appear later. The first will change the eccentricity of comet 2019Y4. The evolution of the orbital elements for both comets is similar over the next 125 kyr. Comet 1844Y1 passed a turbulent orbit evolution unlike comet 2019Y4, which it does not undergo significant changes in the orbital elements and, therefore, it is under the weaker influence of the gravitational forces.

We can deduct from the evolution of the orbital elements that the similarity of the orbits of both comets is significant for at least 180 kyr when the orbits are almost identical. According to our dynamical investigation of the orbits of comets 1844Y1 and 2019Y4, we claim that both comets had no common progenitor because comet 1844Y1 has passed from the hyperbolic orbit to the ecliptic one.

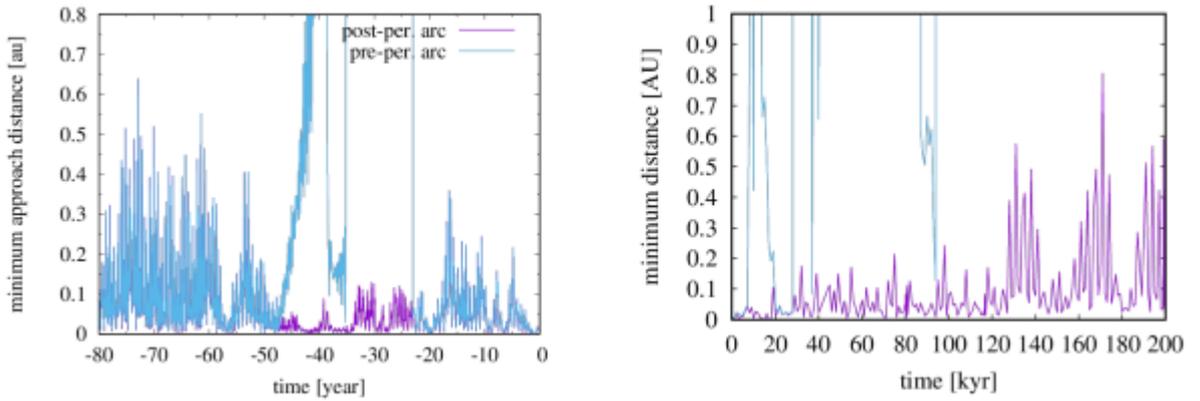

**Figure 11.** Evolution of the minimum distance between the orbital arcs of comet C/1844 Y1 (Great comet) and comet C/2019 Y4 (ATLAS) from 120 kyr in the past up to the present (the left-hand figure) and from the present to 200 kyr in the future. The minimum distance of the post-perihelion arc is shown by the violet curve and that of the pre-perihelion — by cyan).

The minimum distances between comets' (1844Y1 and 2019Y4) orbital arcs during the whole investigated period are shown in Fig.11. The figure on the left side represents the evolution of the minimum distance in the past and the figure on the right side represents the evolution of the minimum distance in the future. The tracking of the evolution ends ~82 kyr in the past when comet 1844Y1 is on the hyperbolic orbit and the minimum distance cannot be checked. The absolute minimum distance between the orbital arcs was 0.0021 au 1.7 kyr ago on the pre-perihelion arc and 0.0034 au on the post-perihelion arc. The maximum distance between the orbital arcs of comets was 33.6867 au 25.5 kyr ago on the pre-perihelion arc. At the time of the predicted split event according to Hui & Ye (2020), the minimum distance between the orbital arcs was 0.017 au. Overall, the mutual distance between the orbital arcs is very short during ~200 kyr (from ~80 kyr in the past to ~120 kyr in the future). The similarity of the orbits implies that both comets have one common progenitor, however, long-term numerical integration reveals that the comets are not genetically related and comet C/1844 Y1 passed from hyperbolic orbit to the elliptic. This is probably due to the close approach to Mercury. The result of numerical integration according to Hui & Ye (2020) is correct within their integration time. Our results on a time scale of ~7 kyr would be identical because the mutual distances of the orbital arcs are small. As well as the evolution of the orbital parameters is similar. That would also lead us to the conclusion about the similarity of comets, respectively about the common progenitor of both comets.

## 6. Summary

We have presented observational data of the split comet 2019Y4 taken with the 6-m Russian telescope on the 14th and 16[th] April 2020. At the time of observations, the heliocentric distance of the comet was 1.212 and 1.174 au, respectively. The geocentric distance of the comet was 0.998 and 0.991 au, and its phase angle was about 52.9° and 54.5°, respectively. We analyzed the images obtained in the cometary filters and the long-slit spectra to derive information on the physical parameters of the cometary coma. We performed the numerical integration of the nominal orbits of comets C/1844 Y1 (Great comet) and C/2019 Y4 (ATLAS) in the past and the future. Dynamical relations between these comets were investigated. Here we resume our main results:

- The contribution of the gas component and the ratio of the emission component to the total flux is calculated. The distribution of the gas component along the slit is asymmetric for solar-anti-solar directions and depends on the orientation of the spectrograph slit. The gas contribution in the anti-solar direction measured in the *BC* filter is negligible, but it fluctuates within 7-8% in the *RC* filter. A very small signal-to-noise ratio measured in the Sun direction results in a large error, nevertheless, the contribution of the gas is quite large in the continuum filter. For the *R* broad-band filter of the Johnson-Cousins system, the contribution of the gas increases from 12% to 18% with the increasing aperture radius. The gas contribution increases up to almost 50% when measured with large apertures in the *B* and *V* filters of the Johnson-Cousins system. The splitting of the nucleus makes this comet quite gaseous.
- The Haser model was used with the purpose to calculate the production rate of gas molecules (Haser, 1957). The $\log[Q_{C2}/Q_{CN}]$ ratios point out that comet C/2019 Y4 (ATLAS) probably belongs to the group of long-period comets according to the dynamic classification (Langland–Shula and Smith, 2011) and to the group of "typical" comets (A'Hearn et al, 1995) according to the compositional classification.
- In the *BC* and *RC* filters, the comet shows an elongated coma along the solar direction and four round condensations (A, B, C, and D) perhaps surrounded by weak small comae. Fragments C and D in the *BC* filter exhibit a much larger size and are shifted relative to the Solar-tail axis. The comet displayed an extended coma with the highly condensed material in the near-nucleus area and a tail in the anti-solar direction over a length near the $1 \times 10^5$ km projected nucleocentric distance. Comparison of the *BC* and *RC* images indicates the similarity of the comet morphology in both filters; although, the coma generated by the dust appears fainter in the *BC* filter than that in the *RC* image.
- Comparing the dust colour and the Sun colour-index reveals the fact that the dust intrinsic colour has changed from moderately red near to the nucleus into immensely blue (the colour-index is equal to approximately 1.0 mag) further away from the nucleus. A significant difference in the behavior of the colour is due to the asymmetry of the gas component's contribution to the total flux along the solar-tail direction. The colour and contribution of the gas component correlate in the Sun direction. In the anti-solar direction, the colour of the coma has a reddish trend from 1.37 to $1.47^m$. The above properties suggest that some evolution of the scattering particles happened, presumably, as a consequence of some composition changes or/and a change in the distribution of particle sizes. In the direction perpendicular to the Sun-tail direction a blueing trend from reddish coma colour, and the asymmetry is observed. Most likely, the asymmetry is related to the contribution of the B fragment.
- Reflectance spectral gradient $S'' \approx 17\%$ per 1000 Å within the wavelength region of 4429 — 6835 Å was calculated directly from the dust particle reflectivity.
- The dynamical evolution of the orbits of comets C/1844 Y1 (Great comet) and C/2019 Y4 (ATLAS) reveals that the Great comet passed from hyperbolic orbit to the ecliptic orbit roughly 82 kyr ago. The orbital transition could have been caused by a close approach of the comet to Mercury at a distance of 0.038 au. Although, from that time, the similarity of the orbits of both comets is significant and such a state will continue for at least 200 kyr. The same is true for the mutual distance between the orbital arcs. During this period the minimum distance between the orbital arcs was 0.0021 au on the pre-perihelion arc and 0.0034 au on the post-perihelion arc. The similarity of the orbits is unquestionable; but from a dynamic point of view, comets C/1844 Y1 and C/2019 Y4 have no common progenitor.

ACKNOWLEDGMENTS

The authors would like to thank the referee Luisa M. Lara and the anonymous referees, whose comments helped to improve the quality of the paper. The observations at the 6-m BTA telescope were financially supported by the Ministry of Education and Science of the Russian Federation (agreement No. 14.619.21.0004, project ID RFMEFI61914X0004). OI thanks the Slovak Academy of Sciences (grant Vega 2/0023/18). This work was supported by the Slovak Research and Development Agency under Contract No. APVV-19-0072. DT thanks the Slovak Grants by the Agency for Science, VEGA, grant No. 2/0037/18, and by the Slovak Research and Development Agency. A grateful acknowledgment is made to Dr. Zubko and Dr. Kaňuchová for the interesting discussion.

DATA AVAILABILITY

The data underlying this article are available in the article and in its online supplementary material.